# History of the NeoClassical Interpretation of Quantum and Relativistic Physics

Shiva Meucci, 2018

ABSTRACT

The need for revolution in modern physics is a well known and often broached subject, however, the precision and success of current models narrows the possible changes to such a great degree that there appears to be no major change possible. We provide herein, the first step toward a possible solution to this paradox via reinterpretation of the conceptual-theoretical framework while still preserving the modern art and tools in an unaltered form. This redivision of concepts and redistribution of the data can revolutionize expectations of new experimental outcomes. This major change within finely tuned constraints is made possible by the fact that numerous mathematically equivalent theories were direct precursors to, and contemporaneous with, the modern interpretations.
In this first of a series of papers, historical investigation of the conceptual lineage of modern theory reveals points of exacting overlap in physical theories which, while now considered cross discipline, originally split from a common source and can be reintegrated as a singular science again. This revival of an older associative hierarchy, combined with modern insights, can open new avenues for investigation. This reintegration of cross-disciplinary theories and tools is defined as the "Neoclassical Interpretation."

INTRODUCTION

The line between classical and modern physics, when examined very closely, is somewhat blurred. Though the label "classical" most often refers to models prior to the switch to the quantum paradigm and the characteristic discrete particle treatments of physics, it may also occasionally be used to refer to macro physics prior to relativity. In the attempt to explicate all the overlapping points of modern and classical conceptualizations, we will discuss the little known alternative fluid dynamical basis of the quantum paradigm, developed alongside the discrete treatment by many of the same scientists, and the interchangeable relationship between discrete and continuous treatments of modern physics. This historical overlap of disciplines will also be revealed to underpin relativistic physics showing that it too can be approached through the lens of hydrodynamics.

It is a less known fact that many modern sciences, such as knot theory and topology, branched from common conceptual roots as the two main branches of theoretical physics. Classical mechanics of the 1700s & 1800s developed for a singular purpose were not abandoned or completely transformed but have branched and continued their own growth separately into the modern age.Through finding modern uses and thus continued development, the interchangeability of the tools and processes across apparently divergent fields has actually been enhanced. The common starting points have led to a direct overlap in these fields and this exchangeability of approach, has inevitably also led to developments in deterministic hydrodynamic analogs of gravity and quantum mechanics which we label "Neoclassical."

To establish the ability to completely exchange conceptual frameworks while maintaining the outcomes of our current successful mechanics, we will explore where and how the analogies between these practices overlap. We will do this through an exploration of the shared historical lineage as well as specific proofs of interchange which have been shown at every point along the history of the shared sciences up to and including the modern day.

**The Parallel History of Hydrodynamics and Electromagnetism**

*"The electromagnetic field behaves as if it were a collection of wheels, pulleys and fluids."*
- James Clerk Maxwell

The early notable developments toward electromagnetic theory were all done in a theoretical environment which presumed a fluid "ether" as the basis upon which all phenomena occurred. In 1746 Euler modeled light in a frictionless compressible fluid.

In 1752 Johann Bernoulli II suggested a model of ether which is a fluid, containing a great number of excessively small vortices. The elasticity of the aether is due to vortices which expand under rotation. A source of light produces perturbation which cause the propagation of oscillations in the ether. Bernoulli compares these oscillations with those of a stretched cord which performs transverse vibrations. (E. Whittaker 1910) Bernoulli's model of ether closely resembles that which was suggested later by Maxwell.

A century later in 1852 Faraday modeled electromagnetism as vibrations in lines of force and it was referred to as "tubes" of force by Maxwell as early as 1855 and by 1861 Maxwell had begun to combine all the prior art into "molecular vortices"; the description of which resulted in the formulation of Maxwell's famous equations.

Most aether theories shared a rotational nature but many difficulties arose from how the substance was treated in an elastic fashion. Maxwell's treatment however, focused more upon the rotational component of the energy and granted the aether qualities which were unfamiliar in fluids and have only more recently been demonstrated to actually exist in superfluids.

Parallel Developmental Paths

While Maxwell's contribution represents a Scottish point of pride which plays a well known role in the development of Special Relativity, a less well known parallel Irish point of equal pride is found in James MacCullagh whose developments prior to Maxwell played an almost unknown, but major role in the development of the theory of relativity.

Beginning with a demonstration of its long reach, MacCullagh's work plays an indirect role in the successful calculation of Einstein's field equations through Max Born's recognition that Gustav Mie's four dimensional continuum could be regarded as a generalization of MacCullagh's three-dimensional aether. Max Born played a primary role in informing David Hilbert on Mie's and Einstein's work and their similarities. (Renn and Stachel 1999) Crucially, during the four weekly publications to the Prussian Academy of Science during November 4-25 of 1915 which are now collectively known as "General Relativity," Einstein was involved in intensive technical and collaborative contact with David Hilbert who was concurrently developing his own attempt at the gravitational field equations using Mie's work as a basis; (Earman & Glymour 1978) though switching the causal hierarchy of electromagnetism and gravity from Mie's theory.

While sources such as Pauli went so far as crediting Hilbert with primacy in the development of the field equations, there is little reasonable doubt that Hilbert and therefore MacCullagh's mechanics, played an influential role in those final weeks and therefore we see some of the first signs of a parallel and analogous mechanics to those that underpin even General Relativity. This demonstration of its far reaching influence will be matched by its very direct role in the development of all those theories that culminated in this modern work.

The defining characteristic of MacCullagh's work on aether is his assertion that the energy of light is not in deformation of the aether but in its rotations. (Renn and Stachel 1999) It is this rotational characteristic that threads MacCullagh's influence all throughout modern theory and most importantly resurfaces at a critical point in Lord Kelvin's work.

Ring Vortex Atoms

In fluid mechanics, Helmholtz's theorems describe the three dimensional motions of fluids around vortex filaments in inviscid flows. In 1867 Lord Kelvin published "On Vortex Atoms" in which he begins by stating that when he first discovered Helmholtz's laws of vortex motion in inviscid fluid, it occurred to him that the ring vortices Helmholtz described must be the only true form of atoms. He then goes on to describe the conservation of these energetic flows and how this idea will fit into the contemporary developments in electromagnetism (Lord Kelvin 1867).

Just one year later in 1868 Kelvin published a more rigorous hydrodynamical description of electromagnetism called "On Vortex Motion" in which he starts:

"the mathematical work of the present paper has been performed to illustrate the hypothesis that space is continuously occupied by an incompressible frictionless liquid, acted on by no force, and that material phenomena of every kind depend solely on motions created in this liquid."

This paper was the beginning of, and the impetus for, the great deal of development in hydrodynamics that Kelvin referred to as "formidable" while Maxwell, in the review of the vortex atom for the 1875 Encyclopedia Britannica called the mathematical difficulties "enormous" though also following that "the glory of surmounting them would be unique."

In 1882 another noteworthy name in the development of the theory J. J. Thomson was awarded the Adams prize at Cambridge for "A Treatise on the Motion of Vortex Rings" in which he was able to show how deformations and configuration of linked rings could provide a mechanical basis for valence (Thomson, J.J. 1883).

Already by By 1880, however, difficulties in treating the elasticity of the medium led Kelvin to postulate a Vortex sponge model that theorists such as FitzGerald and Hicks continued to work with during the mid 1880s. On the apparent necessity of this work Hicks pointed out "The simple incompressible fluid necessary on the vortex atom theory is quite incapable of transmitting vibrations similar to those of light" (Hicks 1885).

The attempt to develop this theory lasted late into the 1880s and the productive work of Kelvin and others eventually led to enormous advances; those advances bore fruit in the form of the discipline we know as hydrodynamics today. This, in turn, paved the way for knot theory and even topology but the discovery of the null Michelson-Morley experiment introduced yet another layer of difficulty to an already long and arduous (but productive) path full of dead ends over nearly a 20 year period.

To recapitulate on the parallel nature of hydrodynamics with relativity, it is useful to note that Einstein remarked in "On the 100th anniversary of Lord Kelvin's birth" on Kelvin's circulation theorem that this development was one of Kelvin's most significant results that provided an early link between inviscid fluid mechanics and topology. (Einstein 1924) We find that nearly all the advanced tools of mathematics used in modern physics have their roots in hydrodynamics.

MacCullaugh's Rediscovery

For instance, in 1878, George FitzGerald discovered that by identifying e with magnetic force, where e is the displacement vector in MacCullagh's theory, and curl e with dialectric displacement, he could obtain the same expressions for kinetic and potential energy in Maxwell's theory as in MacCullagh's, which made MacCullagh's theory of reflection and refraction of light correct in the electromagnetic field free of charges and conduction currents. (Excerpt: Ivor Grattan-Guinness 2002)

In 1880, George FitzGerald repopularized MacCullagh's ideas on a rotationally elastic medium and was able to provide a more solid grounding for the -then faltering- theory of Maxwell. (FitzGerald 1880) What is unclear is why it took roughly a decade for Kelvin to use this same solution of MacCullagh's aether to resolve his own problems with elasticity - shared by Maxwell's theory - which were the major barrier for the vortex atom theory. However, in 1890, less than three years after the Michelson-Morley experiment, Lord Kelvin demonstrated that MacCullagh's aether could be physically realized as a workable mechanical system which would transmit transverse but not longitudinal waves and describe behaviors of aether in precisely the way necessary to account for all known electromagnetic phenomena previous to the MM. (Thomson 1890) This proof of mechanical feasibility runs directly counter to modern claims that

MacCullagh's work was an example of non-physical field ideations which dominate physics post relativity.

The interesting crossover point of note is that George FitzGerald whose work was strongly focussed upon development of Kelvin's vortex atom theory, is well known for collaborating with Hendrik Lorentz on the hypothesis of physical contraction as an explanation for the Michelson-Morley experiment.

Finally, however, the alacrity with which Kelvin finally solves various elasticity troubles of an aether simultaneously capable of transmitting light *as well as supporting his ring vortex model* is shown not only in his article "On a Gyrostatic Adynamic Constitution for Ether" but throughout his volume three of "Mathematical and Physical Papers." It is this re-injection of a workable solution from MacCullagh which will continue to prove the value of a hydrodynamical view of electromagnetism into the modern age. Kelvin's ability to prove the mechanical feasibility and physicality of MacCullagh's aether leads seamlessly and directly to the very first advances toward relativity which were made by Larmor.

Larmor, directly using Lord Kelvin's gyrostatic explication of MacCullagh's aether was able to publish what is now regarded as the "Lorentz transformations" two years before Lorentz, albeit less generally expressed. (Larmor 1897) Lamour closely collaborated with FitzGerald (Buchwald 1995) and his work is well known as one of the forerunners of special relativity.

In this way MacCullagh's work, proven physically feasible by Kelvin, played a irreplaceable role in both special and general relativity while providing a solid mechanical grounding for a hydrodynamical viewpoint which can support both the transmission of light as well as the many hydrodynamical solutions to subatomic phenomena and mechanisms such as valence which J. J. Thomson had made great strides in showing.

The Crucial Deviation Point of Minkowski-Einstein Spacetime

So far we have discussed the interchangeable nature of hydrodynamics with special and general relativity but the transition to the use of the Minkowski convention of four-dimensional spacetime and the attempts of Mie and subsequently Hilbert to expand MacCullagh's work into a four dimensional paradigm seems to suggest a possible quantitative advancement beyond the ultimately three dimensional mechanics of hydrodynamics. This, however, is only a qualitative change and the use of additional dimensions is common in hydrodynamics. The fourth dimension used by both Larmor and Lorentz still described an ultimately three-dimensional reality.

It is widely accepted that Lorentz and Einstein's theories are "mathematically equivalent." It is at this point, however, that a deviation in interpretation leads to a direct change in the outcomes of certain physical considerations. An understanding of the hydrodynamic theories of Larmor and

Lorentz leads to the conclusion that light's constancy is a perspective illusion caused by the very complex interaction of shortening and propagation delays. Therefore any idea of relative simultaneity is also an illusion which is simply mathematical in nature but also present in these precursor theories. In a three-dimensional world, simultaneity is absolute while perception of time is a subjective effect related to light's propagation.

Minkowski, however, was the first to devise a mathematical convention that separates the two *interpretations* and represents constancy and relative simultaneity via the conjoinment of space and time into spacetime. Initially Einstein was wary of Minkowski's convention, and while not rejecting it outright, did show some reluctance at the time in agreeing with the physical meaning it implies which is now the commonplace understanding of the theory.

When Minkowski built up around special relativity a system of "world geometry" that reified the 4th dimension, Einstein initially remarked, "Since the mathematicians have invaded the theory of relativity I do not understand it myself any more" (Sommerfield 1949).

It crucial to note that the hydrodynamics based theories of Larmor and Lorentz also included a fourth dimension and time effects, but not a geometric conjoinment of time with space. The Minkowski convention is a radically different meaning and relationship for the fourth dimension. Just as we classically thought of reality as 3-dimensional but a 4th implied dimension of time was required to represent the whole thing in physics, the reification of the 4th dimension into a physical reality by the Minkowski convention actually implies the possible need for a 5th dimension to encompass the whole of reality, over time, such as proposed by Kaluza-Klein in the early 1920s.

*The mechanics which previously describe an illusion in aether are discarded while the mathematics those mechanics produced are directly imported and interpreted to describe a new arrangement of reality in which the illusion described is, instead, a new truth.*

Because of Einstein's continual reluctance toward Minkowski's convention and thus the physical interpretation inherent therein, the role Minkowski spacetime played in general relativity, may be less than most authors initially suspect. In 1912 after the publication of a paper on March 20th (Einstein 1912a), Einstein was criticised by Max Abraham specifically for not yet using the Minkowski convention.

*"Already a year ago, A. Einstein has given up the essential postulate of the constancy of the speed of light by accepting the effect of the gravitational potential on the speed of light, in his earlier theory; in a recently published work the requirement of the invariance of the equations of motion under Lorentz's transformations also falls, and this gives the death blow to the theory of relativity."* (Abraham, Max 1912)

Nordström who was also of the Minkowski school of thought said, "Einstein's hypothesis that the speed of light c depends upon the gravitational potential leads to considerable difficulties for the principle of relativity, as the discussion between Einstein and Abraham shows us" (Nordström 1912).

On July 4th 1912, Einstein attempted to explain that one must consider the limits of the two major principles, equivalence and of the constancy of light. Further explaining that the constancy of light can only be maintained in spatio-temporal regions of constant gravitational potential. He continues:

*"This is, in my opinion, not the limit of validity of the principle of relativity, but is that of the constancy of the velocity of light, and thus of our current theory of relativity."* (Einstein 1912b)

Thus Einstein clearly outlined that constancy, as it exists in special relativity, is quite different from its expression in a more generalized form within general relativity, therefore we can further see that conjoining space and time may not play any influential role whatsoever in the development of general relativity.

Minkowski spacetime is a methodology of arranging mathematical consideration which is determined by philosophical claims. The difference added by the Minkowski convention, however, is a representation and assumption of the constancy of light and the relativity of simultaneity. This metaphysical or ontological claim, converted into a mathematical methodology can lead to applications of the theory which can lead to specialized hidden outcomes that differentiate the theories which will be discussed in a later paper. The Minkowski convention will be a crucial point of deviation to address in any hydrodynamic recasting of relativity theory.

While the excitement surrounding Einstein's relativity pushed hydrodynamics to the background and the non-physicality of field concepts, plus waves without media, were revolutionary new views of reality, many eminent scientists still continued to favor a mechanical viewpoint as the basis for theory building even long after relativity, and though the apparent connection between hydrodynamics and electromagnetism was beginning to diverge in the mainstream, the equivalence of the two had not been removed.

In 1931 the 73 year old J. J. Thomson, mentor to Ernest Rutherford, still wrote to Lodge that he saw a "close connection between electricity and vortex motion." and further stated, "I have always pictured a line of electric force as a vortex filament'' (Kragh 2002).

Cosserat Continua

The brothers François and Eugène Cosserat represent the final chapter of classical aether theory and the height of the hydrodynamics descending from MacCullagh through Lord Kelvin and also crucially represent a lost history of physics only now truly beginning to resurface. Their work was plagued by the death of - the primary author - François in 1914 and the political environment of science contemporary to their seminal work "Théorie des corps déformables" in

1909 which was, for all intents and purposes, an aether theory which, in a strange repetition of MacCullagh, left it almost utterly unnoticed.

In 1972 around the time the Cosserat brothers were being rediscovered throughout materials science, H. Minagawa wrote that their intent was to complete aether theory and that, they united Maxwell's theory with MacCullagh and Kelvin's work and integrated them into their own theories. (Minagawa H. 1972)

All of our considerations heretofore may be applied just the same to material media as to various ethereal media. We have declared the word matter to be invalid, and what we expose is, as we said to begin with, a theory of action for extension and movement. To have a more complete idea of the notion of matter, we shall explain later on how one must approach the latter from the concept of entropy according to the profound viewpoint that Lippmann introduced into electricity. Excerpt: "Theory of Deformable Bodies," 1909, E. and F. Cosserat, Footnote 1, Chapter 4, section 57

While the work of the Cosserats is widely known to have been ignored by the physics community at large until the 60s and 70s, in the recent book "Wave Dynamics of Generalized Continua" the authors point out the influence of the Cosserats upon contemporary French physicists, Poincare, Picard and Cartan, stating that Cartan's familiarity with the Cosserat work was a major influence in his own future influential work:

"Acquaintance with Cosserat work helped him, the future author of the classic book 'The Theory of Spinors', to create the theory of spaces with torsion. He found a prompting exactly in the Cosserat works: to relate the torsion tensor with internal rotational degrees of freedom of continuous media. In its turn it allowed relating the torsion in space and time with the specific properties of material systems, namely, spin. This connection is achieved in the framework of the dynamic theory of Einstein-Cartan gravitational interactions [81, 145]" (Excerpt: Bagdoev Alexander G. et al 2016, p.226)

This continues the theme of a hydrodynamical aether theory underlying and directly stimulating the development of modern physics at every point along the way, while simultaneously being directly ignored or denigrated.

The Cosserat work, however, is a mechanical and physical theory of electromagnetism which is compatible with Maxwell's electromagnetism, Kelvin's ring vortex atoms, and Poincare's relativistic electromagnetism. Additionally, as we shall explore in a subsequent paper, this interpretation may provide (after much more work is completed) a new mechanical perspective for future science to embrace and thereby possibly realize Einstein and Hilbert's unfulfilled dreams of a unified field theory.

Following a Single Branch to the End.

Given that the Cosserat work is represented herein as the pinnacle of aether theory, it is important to briefly mention the eventual endpoint of this discussion which will progress historically and conceptually beyond the shared history of particle theory and hydrodynamics discussed in the next section. While this breaks the historical flow, the contextual continuity is useful in conveying the overarching theme of the interpretation described by this paper.

In the late sixties, as the rediscovery of the value of the Cosserat work was beginning to come to the fore, recognition of this system as a description of spacetime led J. A. Brinkman of the very notable aerospace company, Rockwell Corporation, to modernize MacCullagh's aether as a model for electromagnetism in the modern context by adapting it to a 4 dimensional model. (Brinkman J.A. 1968) More recently, however, it's been shown that there are various multi-dimensional derivations of Cosserat continua which can support covariance (Panicaud, B. & Rouhaud, E. 2016) and thus provide better analogy for relativistic spacetime.

Specifically, the endpoint of significant note is that the inviscid fluid medium described, from MacCullagh and Maxwell up through Kevlin to the Cosserats, has most of the properties now known as a superfluid. The exploration of superfluid models of spacetime is not a new concept with some of the first major papers published circa 1976, but the Cosserat model of precisely how a superfluid vacuum might work is significantly different from any previous proposals with numerous implications for the application and interpretation of Lorentz invariance as well as expectations about the phenomena of vortex filaments proposed in those earlier theories and now shown to exist via experiments with superfluid helium. (Donnelly RJ 1991)

While at least one author has demonstrated a "precise analogy between superfluids and Cosserat continua" in the context of relativity, (G. Ferrarese 1997) there have also been other recent advancements in applying micropolar elasticity, originally developed by the Cosserats, to relativistic field theories. (Bohmer, CG; Obukhov, YN; 2012)

These modern proofs of the close agreement of the Cosserat brothers' model of elasticity with current relativistic physics and indications of further advancements to be found using this methodology can distract from the fact that the physics being used is the physics of an aether theory, or more crucially hydrodynamics, not simply a special form of elasticity.

It is useful to bear this in mind as we take a step back to examine the historical role hydrodynamics has played in the development of quantum theory.

**The Hydrodynamic Analog in Quantum Physics**

As early as 1926 just one year after the formulation of Shroedinger's equation, Erwin Madelung demonstrated that it can be recast in hydrodynamic form. The Madelung equations are Eulerian and very directly show the relationship between fluid dynamics and quantum mechanics (Madelung 1926,1927).

By 1950 the phase space formulation of QM provided a fully realized and valid formulation of the science which could deal with both position and momentum in a classical manner. The Wigner-Weyl transform provides a mapping between this formulation of QM and the Hilbert space operators in the Schroedinger representation which is completely equivalent (Curtright et al 2005, 2012). This shows the direct mathematical compatibility of more classical treatments (Hamiltonian) and quantum mechanics. Additionally, it has been shown that the Wigner quasiprobability distribution function can, via deformation, match a phase space distribution which describes de Broglie-Bohm causal trajectories (Dias, Prata 2002).

These relationships show how the deterministic chaos found in hydrodynamic systems can lie underneath the probability distributions used in quantum mechanics and thus continues the theme of interchangeability between hydrodynamics and all aspects of modern theory.

A Word on Transitioning from Discrete to Continuous

Phonons are an example of applying the methods of quantum mechanics to the treatment of mechanical waves in materials in which the continuous phenomena of conventional mechanical waves, can be treated as discrete particles with all the behaviors and expectations of any other quantum mechanical particles. These quasiparticles, however, have no physical existence apart from the medium, but their correspondence to border conditions and behaviors of the material make them an extremely useful convention. This use of quantum mechanics to solve problems within the continuum mechanics domain shows a well known direct interchangeability of the systems in one direction. The only constraint to going in the other direction is tradition and the modern question of if chaotic determinism truly underlies the probability found in quantum mechanics.

The most important lesson to gather from phonons is that "particles" can represent border transitions and other real world phenomena in an abstract fashion, but the discrete nature of the particle is an illusion of the mathematical treatment. The usefulness of the abstraction and its direct relationship to real-world effects can lead to a false intuition that the particles exist as discrete self-contained entities in physical reality when they factually do not. They are partitions for characteristics. Collective actions of groups appeal to human sensibilities because of their everyday use. We say "congress voted," or "the crime syndicate pressured businesses" and even if the individuals of these two groups overlap, it is common to think of these entities as separate extant objects when they are, in fact, separate but overlapping *interpretations* of data.

In the theoretical physics world, the reification of waves is deeply ingrained as a cultural normative thought pattern which sometimes opposes the recognition of waves as collective actions of physically extant substance. Thus superposed mechanical waves might be thought of as separate entities, but with a moment's consideration, they can be recognized as actually overlapping categorizations and not factually overlapping objects. As we will show, this understanding will also apply to other particulate treatments in physics under a hydrodynamic interpretation.

The Unnecessary Micro-Macro Divide

In a little over the past decade, all of the strange and seemingly non-physical behaviors of quantum mechanics, which were previously believed impossible in the macro world, have been replicated in numerous fluid mechanics experiments. This includes the dual-slit experiment, orbital quantization, single-particle diffraction, zeeman splitting, quantum tunneling and a host of other behaviors typically only associated with the mechanics far below the observable macro realm.(Couder et al 2005)(Fort & Couder 2006)(Eddi et al 2009)(Eddi et al 2012)

The first of these experiments were performed by Yves Couder who used a vibrating silicon oil bath to create a situation in which a millimeter-sized droplet bounces indefinitely upon the surface. These "walkers" move along the surface interacting with the waves they produce. The location of a walker confined to a cavity and tracked over time, can be described as a probability distribution which is the Faraday wave mode of the cavity, just like a particle in a quantum corral.

The similarity of this system with quantum mechanics is best described by a leading experimental researcher at MIT in the August 2015 issue of Physics Today:

"The walker represents an example of an oscillating particle moving in resonance with its own wave field. The droplet moves in a state of energetic equilibrium with the vibrating bath, navigating a wave field sculpted by its motion. The walker system continues to extend the range of classical systems to include features previously thought to be exclusive to the quantum realm. What might one infer if unaware that it is a driven, dissipative pilot-wave system? One would be puzzled by the prevalence of quantization and multimodal statistics. Inferring a consistent trajectory equation would be possible only in certain limits. Doing so in the limit of weak walker acceleration would suggest that the droplet's effective mass depends on its speed. Multiple particle interactions would be characterized by inexplicable scattering events and bound states, and baffling correlations. If one could detect a walker only by interacting with the fluid bath, the measurement process would become intrusive. If a detector confined the walker spatially, one would infer a position−momentum uncertainty relation. If detection required collisions with other droplets, disruption of the pilot wave would destroy any coherent statistical behavior that might otherwise arise." - Excerpt: Physics Today, August 2015, John M.W. Bush

Modern Examples of Neoclassical Theory

With so many mechanical explanations for quantum mechanical behaviors provided by the walker experiments, it becomes necessary to examine any behaviors of quantum mechanics which are not described by these experiments. The apparent differentiators of first interest are the non-local effects found in entanglement and the physical interpretation of quantum uncertainty as ontological indeterminacy instead of complex chaotic determinacy.

It must first be recognized that the wave of probability which can describe a walker is separate from its pilot wave. This system is described by Bell as a close approximation of de Broglie's later "double-solution theory" in which de Broglie stressed the "harmony of phases" and is shown in the experiments by the statistical wave and the pilot wave having the same wavelength but different geometric form.

In the analogy provided by the walker experiments, the dynamics of the system can be described as a non-local. In this case, the complex state of the system at any point point in time describes a non-local hidden variable theory. These experimental findings in combination with the interchangeability between quantum mechanics and fluid dynamics has led authors from previously disparate fields to be able to propose deterministic dynamics which shed light on a possible mechanics underlying seemingly random behaviors. The differentiation between particle and quasiparticle is a primary area of exploration for the Neoclassical Interpretation.

Most notably, Ross Anderson and Robert Brady of Cambridge have made major contributions to the exploration of this interpretation as well as solid theoretical grounds for deeper investigations. In "Maxwell's fluid model of magnetism" the authors show that a wavepacket travelling along a phase vortex in an Eulerian fluid obeys Maxwell's equations, is emitted and absorbed discretely, and can have linear or circular polarisation. Additionally, the measured correlation between the polarisation of two cogenerated wavepackets is exactly the same as predicted by quantum mechanics, and observed in the Bell tests. (Brady, Anderson 2015). Another paper by the group discusses a violation of Bell's inequality in fluid mechanics thereby providing underlying deterministic mechanics in place of the arguably less physically real models of the behavior (Brady, Anderson 2013). Robert Brady in "The Irrotational motion of Compressible Inviscid Fluid" outlines a comprehensive physical mapping between quantum mechanical theory and fluid dynamics in which he not only provides an analog for gravity but also introduces the concept of a relativistic quasiparticle called a "sonon" which exhibits spin ½ symmetry (Brady 2013). This sonon concept bears a striking resemblance to the smoke ring or "Vortex Atoms" initially proposed by Lord Kelvin, thus further continuing the neoclassical theme to this interpretation (Silliman 1963).

At the intersection of the various application of continuum mechanics and electromagnetism, another scientist, David Delphenich, who also translated the Cosserat's 1909 book to english, has spent many years researching the most finite details of a neoclassical approach to physics with deep dives into the geometry and topology of spacetime and independently arrived to the nomenclature or label "neoclassical" for his work. Hence the "Neoclassical Interpretation" currently exists as disparate scientists and researchers coming to nearly identical and compatible conclusions which are informed by the hydrodynamical roots of electromagnetism.

DISCUSSION

What Advantage is Equivalent Alternative?

Given that the current quantum mechanical paradigm has been very successful and the same is true of relativistic physics, the question that occurs is what advantage might the use of a different interpretation and set of mechanics have if they are already completely interchangeable? The idea of quasiparticles being simply properties of a substance, however, provides for the possibility of direct manipulation of the medium, and analogies to the manipulations of fluids, when applied to the vacuum, may lead to entirely new experimental processes. Completely new experimental regimes can be embarked upon in the investigative process.

From the relativistic perspective, it is often improperly assumed that a medium for electromagnetic effects is mutually exclusive with the principle of relativity. A conflict, however, is only true for the presumption of one-way light speed isotropy in relativity, which has never been experimentally verified. The Lorentz transform, however, remains valid in preferred frame mechanics in which the one-way speed of light is anisotropic because the two-way speed of light can remain isotropic under the same circumstances.

In 1920 at the university of Leiden, Einstein extolled the virtue and absolute necessity of an "aether" and furthermore praised Mach's idea of a gravitationally mediated preferred frame derived from the average gravitation of the universe, but Einstein still resisted a preferred frame mechanic for electromagnetism. Lorentz invariance, however, is always maintained in two-way tests of light speed. Therefore the possibility of preferred frame mechanics still remains fully within the bounds of modern physics and also rests under the umbrella of a "Neoclassical Interpretation."

*"Recapitulating, we may say that according to the general theory of relativity space is endowed with physical qualities; in this sense, therefore, there exists an ether. According to the general theory of relativity space without ether is unthinkable; for in such space there not only would be no propagation of light, but also no possibility of existence for standards of space and time (measuring-rods and clocks), nor therefore any space-time intervals in the physical sense."* - Albert Einstein, University of Leiden 1920

Sociology of Scientific Revolution

While these explorations have opened a new field of inquiry ripe for further study, the re-introduction of determinism as a possible underlying mechanic for the universe may leave investigations of consciousness and spirituality in a very uncomfortable existential position and, on the surface, seemingly a scientifically untenable one. The removal of pure probability as a basic source of reality simultaneously removes any magical or non-mechanical explanation of spirituality. The illusory nature of constancy in relativity, completely removes relativity as a core precept of the more fuzzy concept of "relativism" that is often applied to philosophical, sociological, and moral discussions.

Even those scientists and researchers who are atheist in philosophy can suffer from hidden biases that stem from the natural human desire for self respect as well as the necessity of individual responsibility inherent in consideration of free will when attempting to create systems of law and governance. The cognitive difficulties of the concept of determinism are far reaching and multi-faceted.

This narrowing of possibilities for spirituality under determinism and a rational set of mechanics seems to, once again, lean strongly toward the pure "materialism" perspective which has been heralded as a dead-end for over half a century. Given the newest developments in information theory, the role of information in concepts of thermodynamics may still lead to the conclusion that a purely "material" idea of the world could still be considered insufficient because states, conformation, and "potentials" and the very concept of "energy" itself may eventually be labeled as very specifically non-material in nature. This sociological and deeply ingrained psychological desire to uphold the freedom of the human mind and the possibility of systems outside our current understanding can be addressed and ameliorated via considerations of the interactions of information with physical reality through information theory. The identification of rational mechanisms for data storage and calculation as "spiritual" in nature may provide a linguistic doorway through which the spiritual and scientific communities may find compatibility never before achieved (Meucci Shiva, 2015).

Defining the Differences

For the purpose of outlining the historical impetus for exploring the topic, this paper has focused upon the similarities and interchangeable concepts and tools that overlap between a Neoclassical Interpretation and the more commonplace interpretations. However, a subsequent paper will focus upon the crucial differences between the predictions and processes found in the prior interpretations and those proposed by Neoclassical Interpretation.

The development of science and especially theory must always rely upon experimentation as the final selection of a superior theory and thus the next paper(s) will focus upon defining characteristics and how those crucial deviations can be tested.

CONCLUSION

A thorough examination of the historical process involved in the development of the modern tools for theoretical physics shows a compatibility between the methods of hydrodynamics, materials science, and the tools for dealing with particle physics, electromagnetism and even gravity.

The conceptual lineage of the Neoclassical Interpretation traces the roots of electromagnetism back to MacCullagh and places his contribution on equivalent, and in some cases superior, footing with Maxwell's contribution and thus projects the future of investigations in theoretical physics may list MacCullagh, Kelvin and the Cosserats as primary fathers of 21st century physics.

A new paradigm is forming just under the surface of modern theoretical physics which will change very little of what has been done but may vastly change our notions of what can and will be done in the field. The tools that most likely will be employed will be a marriage of hydrodynamics and materials science pioneered by the Cosserat brothers.

It could provide a rational basis of mechanics for all phenomena which will reinstate a deterministic view of reality that will allow the enforcement of a "no magic" rule in science once again, and therefore re-institute a mechanical requirement for all proposed phenomena in science thus eliminating any variety of "spooky action at a distance" or conflation of properties of objects as physically extant and separate objects. This clearer re-division of actual particles from virtual particles may allow further re-partitioning of subjects which might grant new insights. Through the language of information theory, this re-partitioning may give greater clarity to the bounds and exchange between that which we call "information" and that which we label as "material."

The Neoclassical Interpretation will expose a new view of material phenomena under Kelvin's concept of vortex filaments and their various configurations. These configurations will give a new-yet-old mechanical basis for atomic structure and valence proposed by JJ Thomson which, by providing a mechanical reasoning for atomic bonds, will expand our ability to propose and explore other mixed or conjoined subtleties in those interactions which may have previously presented themselves in prior art as one of the various and prolific virtual particles.

In this framework, concepts of probabilistic methods will be separated from extant phenomena and conceived of as simply necessary tools for the arbitrary partitioning of continuous systems that behave via chaotic determinism.

Additionally, this new view of the role of chaos theory and chaotic determinism underlying physics may even lead towards a new view of "cause-and-effect" that recognizes the fundamentally misleading nature of a (timewise) linear focus upon causes of events. The lateral, or "simultaneous," nature of constraints and influences that determine events are equally causal in a given event when we consider the complex contextual dependencies found in deterministic chaos, such as observed in the "walker" experiments. Common views of determinism in the past were almost exclusively focussed upon linear prior chains of events, but chaotic determinism refocuses our concept of "causal events" upon the instantaneous configuration of the universe as crucial to the nature of those events. Thus, while these factors were not unconsidered, causality may begin to be viewed as possessing at least two equal contributions comprised of the moments from the past as well as a dramatically increased importance of the configuration of the instant under consideration.

The exploration of overlap between particles and properties will be a fertile field to understand the many virtual particles which can be re-identified and re-partitioned as behaviors and properties of a media which will be treated as continuous down to and until near the Planck length.

Revolution in science is constrained by enormous bodies of evidence and as technology increases, the errors of science are pushed into the realm of human cognition, expectation, bias and, most crucially, the interpretation of the data. It is the vast and important work of all those who have contributed to the refinement of the standard model that have brought it to its current breaking point. The continual breakdown of the standard model and the various crises such as the vacuum energy catastrophe, as well as model inconsistencies between relativistic theory and quantum theory are a valuable fruit of scientific work which are of inestimable value: the identification of a critical need for revolution.

The finite nature of the constraints we have developed, however, lead to an illusion of perfection which prevents any revolution from being possible. Therefore some finite constraints must be lifted in the pursuit of the error they have identified. The enormous intensity of the errors we have identified lead to conclusion of a need for fundamental revolution while the fantastic success of our methods leads to utterly the opposite conclusion. This dichotomy leads to a nearly impossible task of identifying which constraints can or should be relaxed.

Only through reworking our current model in a different form and then coming to the very same conclusions through a new route, can we differentiate the finite constraints that are actually in error from those that are still necessary. The only solution to the dichotomy of a need for fundamental revolution within an enormous body of data and successful processes, is a revolution of interpretation which can occur through multi-disciplinary approaches. The Neoclassical Interpretation is that revolution which is already occurring but simply needs a greater level of organization and communication between its various parts and participants so that it can garner the necessary funding and community support to complete the enormous project of recasting the whole of the current picture in a different form.